# Adaptation and Validation of the Turkish Version of the Large Language Model Dependency Scale (LLM-D12)

Tuğba Coşkun Aslan[1], Gülser Uncular[1]*, Hasan Durmuş[1], Yasin Kavla[2], Arda Borlu[1], Sameha Alshakhsi[3], Ala Yankouskaya[4], Raian Ali[3]*

[1]Department of Public Health, Erciyes University, Kayseri, Türkiye
[2]Department of Psychiatry, Cerrahpaşa Medical Faculty, Istanbul University-Cerrahpaşa, Istanbul, Türkiye
[3]College of Science and Engineering, Hamad Bin Khalifa University, Doha, Qatar
[4] Department of Psychology, Bournemouth University, Poole, United Kingdom

**Corresponding authors**

*Gülser Uncular, Department of Public Health, Erciyes University, Kayseri, Türkiye. Email: gulseruncular@erciyes.edu.tr*

*Raian Ali, College of Science and Engineering, Hamad Bin Khalifa University, Doha, Qatar. Email: raali2@hbku.edu.qa*

**Abstract**

This study aimed to adapt the Dual-Dimensional Scale of Instrumental and Relational Dependencies on Large Language Models (LLM-D12) into Turkish and evaluate its psychometric properties among regular LLM users. A sample of 387 participants (68.5% female; mean age = 25.22 ± 7.13) completed the translated scale, which underwent cultural-linguistic validation through forward–backward translation and expert review. Confirmatory factor analysis supported the original two-factor structure after removing one item, with strong model fit (CFI = 0.993, RMSEA = 0.073). Internal consistency was high across both subscales: Cronbach's alpha = 0.831 (instrumental), 0.876 (relational), and 0.868 (total); McDonald's omega = 0.834, 0.880, and 0.900, respectively. Test–retest reliability and item monotonicity were satisfactory. External validity was demonstrated via significant associations with ATAI, IA, and PTLLM scores. Interestingly, the lack of association with need for cognition (NFC) suggests that LLM dependency may reflect strategic cognitive offloading rather than cognitive avoidance. The Turkish version of the LLM-D12 is a valid and reliable 11-item tool for assessing both instrumental and relational dependencies on LLMs.

# INTRODUCTION

Large language models (LLMs) have experienced growing use in both personal and professional settings over the past few years [1]. These artificial intelligence (AI) systems, underpinned by deep learning techniques and trained on expansive datasets, exhibit advanced capabilities in interpreting and generating human language, facilitating a diverse range of language-based activities [2].

In Turkey, enthusiasm for and integration of AI technologies have accelerated markedly. Artificial Intelligence Statistics for 2025 report that enterprise adoption of at least one AI technology advanced from 2.7% in 2021 to 7.5% in 2025. Among individuals aged 16–74, 19.2% indicated AI utilization, with pronounced uptake among younger cohorts (39.4% for ages 16–24; 30.0% for ages 25–34) and those possessing tertiary education (36.1%). AI is chiefly deployed for personal ends, including information seeking and entertainment, alongside professional applications (33.8%) and formal learning (31.4%) [3].

With LLMs embedding further into routine practices, they have transitioned from informational resources to interactive social entities adept at empathetic communication, cultivating user confidence, and forging affective ties [4]. Platforms such as OpenAI's ChatGPT and Google's Gemini confer notable pragmatic advantages—task acceleration, expeditious information retrieval, and comprehensive aid—prompting overreliance through their proven performance and dependability [5]. The perceived empathy and non-judgmental nature of AI systems, combined with their accessibility, have led users to develop a problematic emotional reliance that often overrides rational privacy concerns [6]. Such trust often develops based on the perceived success and consistency with which AI performs assigned tasks [7]. As individuals increasingly delegate reasoning and decision-making processes to these technologies, dependency may emerge. When users start to rely on AI systems habitually, often without critical reflection, this dependency can contribute to a decline in critical thinking, creativity, and independent reasoning skills [8]. Recent empirical studies have demonstrated that user dependency on generative AI extends beyond mere information seeking, often serving needs for emotional regulation and social connection [9]. This psychological shift has led to the development of specialized measurement tools, such as the LLM-D12 scale, to assess both instrumental and relational dimensions of this dependency [5].

Dependence on LLMs should be addressed not only at the functional level (e.g., reliance on artificial intelligence for routine tasks such as drafting emails or translating text) but also at the socio-emotional level (e.g., expectations of comfort, emotional support, social approval, or the formation of friendships) [10]. The existing literature includes a limited number of studies that consider both levels simultaneously and assess dependency potential in a multidimensional manner. The Large Language Model Dependency Scale (LLM-D12) was developed to evaluate problematic trust and dependency patterns specific to LLMs and is a measurement instrument consisting of two sub-dimensions, "instrumental dependency" and "relational dependency," which are conceptually distinct from other behavioral dependency scales [5]. Instrumental dependency refers to excessive reliance on LLMs for task completion, problem-solving, and decision-making processes. In contrast, relational dependency encompasses socio-emotional tendencies such as seeking emotional support, perceived social bonds, and friendship [5].

Despite the rapid proliferation of LLM use in Turkey, particularly among young and highly educated groups who intensively utilize these technologies for specific purposes, there are very few studies examining the risks of instrumental and relational dependency on LLMs within the context of Turkish society. This gap underscores the need for a comprehensive assessment of both instrumental and relational dependency risks associated with LLM use in Turkey. The LLM-D12 scale has been developed and validated in English-speaking populations [5]. Accordingly, the present study aims to adapt the LLM-D12 into Turkish and to reliably assess the instrumental and relational dimensions of LLM dependency within Turkish population.

**METHOD**

**Sample:** The population for this methodological study consisted of individuals aged 18 years and older residing in Turkey. Data were collected between September and October 2025, and a total of 387 individuals who agreed to participate were included in the study. Given that LLM usage requires a certain level of digital literacy and technological access, the sampling strategy prioritized individuals with higher education levels to ensure data relevance and reliability. Individuals with physical or psychiatric conditions that could impede communication, as well as those unable to read or write in Turkish, were excluded from participation. Inclusion criteria required participants to use at least one LLM website or application, such as ChatGPT (OpenAI), Gemini (Google), Claude (Anthropic), Copilot (Microsoft), DeepSeek (DeepSeek Inc.), Ernie (Baidu), Llama (Meta), Grok (xAI), or Mistral, daily or to report a meaningful level of reliance on these technologies, and to identify themselves as Turkish in terms of culture and norms. In validity and reliability studies, a commonly accepted guideline is to include at least 10 participants per item for factor analysis [11, 12]. As the questionnaire consisted of 12 items loading on two latent factors, sample size requirements were guided by established recommendations for factor analytic validation studies. A commonly cited rule of thumb suggests a minimum ratio of 10 participants per item to obtain stable factor solutions, which corresponds to a minimum sample size of 120 participants [13]. However, larger samples are recommended to improve the stability and replicability of factor loadings, increase statistical power, and permit more reliable estimation of model parameters, particularly in confirmatory factor analysis [14]. Therefore, the study aimed to substantially exceed the minimum requirement and was completed with 387 participants. Prior to the study, all participants were briefed on the research objectives and procedures, and written informed consent was obtained from each volunteer. The research questionnaires were administered under the supervision of the researchers. To assess the reliability of the measurement instrument, the questionnaire was re-administered one week after the initial application to 55 participants randomly selected from those who had completed the first survey.

**Measures:** The study data consisted of sociodemographic questions developed by the researchers; items assessing LLM usage status and Perceived Trustworthiness of primary LLM (PT-LLM-08), A Dual-Dimensional Scale of Instrumental and Relational Dependencies on Large Language Models (LLM-D12), Attitudes Toward AI (ATAI), Internet Addiction (IA), Need for Cognition (NFC), and attention checks.

***The Dual-Dimensional Scale of Instrumental and Relational Dependencies on Large Language Models (LLM-D12):*** The instrument comprises two subscales and 12 items, rated on a 6-point Likert scale: Instrumental Dependency (6 items) and Relational Dependency (6 items) (Table 1). Response options range from 1 (Strongly Disagree) to 6 (Strongly Agree). The Cronbach's alpha coefficient for the Instrumental Dependency subscale is 0.84, and for the Relational Dependency subscale is 0.91. Instrumental dependency assesses habitual reliance on LLMs for memory, decision-making, task automation, and the simplification of cognitive processes, capturing the transfer of cognitive load to LLMs and the resulting tendency toward continuous use. Relational dependency measures users' emotional and social attachment to LLMs and, based on anthropomorphism and parasocial interaction, encompasses attachment patterns involving the perception of LLMs as companions, the attribution of human characteristics to them, and a reduced desire to communicate with humans [5]. The Turkish version of the survey is presented in Appendix 1 (available at https://osf.io/r978t).

Table 1. Mapping of the Original LLM-D12 Items and the Adapted Turkish Version

| Item | Original Item | Turkish Item |
|---|---|---|
| ID1 | Without it, I feel less confident when making decisions. | O olmadan karar verirken kendime olan güvenim azalıyor. |
| ID2 | I use it sometimes without realizing how much time I spend immersed in it. | Ne kadar zaman geçirdiğimin farkına varmadan bazen onu kullanıyorum. |
| ID3 | I feel much more rewarded and pleased when completing tasks using it. | Görevleri onunla tamamladığımda çok daha tatmin olmuş ve memnun hissediyorum. |
| ID4 | I turn to it for support in decisions, even when I can make them myself with some effort. | Biraz çaba göstererek kendim de verebileceğim kararlar için bile ona başvurduğum oluyor. |
| ID5 | It is my go-to for assistance in decision-making. | Karar verme süreçlerinde ilk başvurduğum destek kaynağı odur. |
| ID6 | Making decisions without it feels somewhat uneasy. | Onu kullanmadan karar vermek beni bir miktar huzursuz hissettiriyor. |
| RD7 | I share details about my private life with it. | Özel hayatıma dair ayrıntıları onunla paylaşıyorum. |
| RD8 | I interact with it as if it were a genuine companion. | Onunla etkileşim kurarken gerçek bir arkadaşmış gibi davranıyorum. |
| RD9 | It helps me feel less alone when I need to talk to someone. | Konuşacak birine ihtiyaç duyduğumda, kendimi daha az yalnız hissetmemi sağlıyor. |
| RD10 | It adds to my social life, making socializing more engaging and interesting. | Sosyal hayatıma katkıda bulunuyor sosyalleşmeyi daha ilgi çekici ve keyifli hâle getiriyor. |
| RD11* | I do not expect it to understand me, nor do I share or discuss my feelings with it. I use it solely as a tool. | Onun beni anlamasını beklemem; onunla duygularımı paylaşmaz ya da tartışmam. Onu yalnızca bir araç olarak kullanırım. |
| RD12 | It helps me feel less alone, reducing the need to talk to others. | Onunla etkileşim kurmak, başkalarıyla konuşma ihtiyacımı azaltarak kendimi daha az yalnız hissetmemi sağlıyor. |

*Note:* **ID** = Instrumental Dependency; **RD** = Relationship Dependency
\* Original Item 11 was removed from the final Turkish scale

***Attitudes Toward AI (ATAI):*** Developed to assess attitudes toward artificial intelligence, the ATAI consists of two subscales: ATAI Acceptance (2 items) and ATAI Fear (3 items). All items are rated on an 11-point Likert scale ranging from 0 (Strongly Disagree) to 10 (Strongly Agree). Total scores range from 0 to 20 for ATAI Acceptance and from 0 to 30 for ATAI Fear [15].

***Internet Addiction (IA):*** The seven-item short form of the Internet Addiction Test (IAT-7), used to assess internet addiction, is scored on a five-point Likert scale ranging from 1 (Never) to 5 (Always). Total scores on the scale range from 7 to 35 [16].

***Need for Cognition (NFC):*** The six-item short form of the Need for Cognition (NFC) scale is scored on a five-point Likert-type scale ranging from 1 (Extremely uncharacteristic of me) to 5 (Extremely characteristic of me). Total scores on the scale range from 6 to 30 [17].

***Perceived Trustworthiness of Primary LLM:*** The participants' perceived level of trust in their primary LLM is assessed using an eight-item scale adapted from a previously developed evaluation framework. The scale encompasses eight core dimensions of LLM reliability and accuracy, security, fairness, robustness, privacy, machine ethics, transparency, and accountability, each represented by a single item. All items are rated on an 11-point Likert-type scale ranging from 0 (Not at all) to 10 (Completely). The total trust score obtainable from the scale ranges from 0 to 80 [18].

*The use of LLMs:* Participants' frequency of LLM use is assessed using an 11-point Likert-type scale ranging from 0 (Very infrequently) to 10 (Very frequently). Data are also collected regarding the mode of interaction with the LLM (text, voice, text and voice, or other), the preferred device for accessing the LLM (computer, mobile device, tablet, smart speaker, dedicated AI device, or other), and the context in which the LLM is most commonly used (stationary, on the move, both).

Attention check items were included in the survey to enhance data quality. These items were designed to be similar in language, structure, and length to the other scale items, thereby reducing their detectability and identifying careless or inattentive responses.

**Language and Content Validity of the Measurement Instrument:** The translation–back translation method was employed in the language validation process of the scales, following the guidelines for cross-cultural adaptation of self-report measures [19, 20]. The scales were independently translated into Turkish by two translators whose native language is Turkish and who are highly proficient in English. The research team reviewed the two Turkish versions and agreed upon a single consolidated Turkish version. This version was then translated back into English by an independent translator who was blind to the original scale. Subsequently, an Expert Committee consisting of 10 specialists in public health and psychiatry was formed. This committee compared the back-translated English version with the original English items to ensure conceptual equivalence and identify any semantic discrepancies. Finally, the Turkish version was consolidated based on the committee's consensus recommendations.

**Analysis:** IBM SPSS Statistics 25.0 software was used to organize the research data and to compute descriptive statistics. Monotonicity, Network, Confirmatory factor analysis (CFA), and reliability analyses were conducted in the R Studio environment (version 4.3.2) to examine the factor structure of the scale. Within the scope of descriptive statistics, the distributional properties of the variables were analysed using skewness and kurtosis values as well as appropriate graphical methods. Given the Likert-type and ordinal nature of the items used in the CFA, estimation methods that do not require the assumption of multivariate normality were preferred. All statistical analyses were performed in accordance with the assumptions of the respective analytical procedures.

**Validity and Reliability:** The scale had previously been developed, and its factor structure was established. Subsequently, we examined the item-level psychometric properties, taking in to account the presence of a reverse-coded item in the Relational Dependency subscale and the potential effects of reverse-coded items on factor structure and internal consistency, as noted in the literature.

*Monotonicity analysis*

The aim of this analysis was to examine whether the items of the instrumental and relational dependency scales satisfied the assumption of monotonicity [21]. Item-level monotonicity was examined using nonparametric item response theory within the Mokken scaling framework [22]. For each item, respondents were grouped into rest-score intervals excluding the target item, and conditional response probabilities were estimated across these score groups.

Prior to CFA, the suitability of the data for factor analysis and sample size adequacy were examined using the Kaiser-Meyer-Olkin (KMO) measure and Bartlett's Test of Sphericity. Following Monotonicity, CFA was conducted to validate the factor structures of the subscales. CFA was performed under the assumption of a single-factor structure for each subscale, and all analyses accounted for the ordinal nature of the multicategorical (Likert-type) items. Within this scope, the Weighted Least Squares Method (Weighted Least Squares Mean and Variance adjusted; WLSMV) was used for model estimation. Factor loadings were reported as standardized coefficients, and each item's contribution to the factor was evaluated based on standardized factor loadings, standard errors (SE), z-statistics, and explained variance ($R^2$) values. Item uniqueness values were calculated to represent the proportion of item variance not explained by the factor. Model fit was evaluated using the Comparative Fit Index (CFI), Tucker–Lewis Index (TLI), Root Mean Square Error of Approximation (RMSEA), and Standardized Root Mean Square Residual (SRMR). The statistical significance of all item factor loadings was assessed using z-statistics, with a significance level of $p < 0.001$. Standardized factor loadings, standard errors, uniqueness values, and explained variance associated with the confirmatory factor analysis were calculated (Table 4).

The *internal consistency and item analysis results* were calculated based on the final set of scale items included in the CFA. Accordingly, internal consistency was assessed using *Cronbach's alpha coefficient* for each subscale. Corrected item-total correlations (r.cor) were calculated to examine item-level performance, along with the change in Cronbach's alpha if an item was deleted (alpha if item deleted), the standard error of alpha (alpha SE), the variance of item-item correlations (var.r) and adjusted item-total correlations obtained when an item was deleted (r.drop) (Table 5). In addition, the reliability of the overall scale and its subscales was evaluated using *McDonald's omega (ω) coefficient*, which addresses limitations of Cronbach's alpha in multidimensional scale structures.

*Network analysis*

The aim of the network analysis was to estimate the pattern of conditional associations among six instrumental dependency items and five relationship dependency items. By modelling instrumental and relationship dependency items within a single network, this approach allows the identification of items that are highly connected within their respective dimensions or that serve as bridges between instrumental and relationship dependency.

A regularised partial correlation network was estimated to model the associations among the dependency items. To account for the ordinal response format, polychoric correlations were computed between all items. The resulting correlation matrix was symmetrised and, where necessary, adjusted to ensure positive definiteness using a nearest positive definite transformation. Network estimation was then performed using the graphical least absolute shrinkage and selection operator (LASSO) with Extended Bayesian Information Criterion (EBIC) model selection. This regularisation approach yields a sparse and interpretable network by shrinking small partial correlations to zero. The EBIC hyperparameter was set to $\gamma = 0.50$.

Edges in the network represent regularised partial correlations between pairs of items, indicating the association between two items after controlling for all other items in the network. Node-level centrality indices were computed from the estimated weight matrix: strength, defined as the sum of absolute edge weights connected to a node, and expected influence, defined as the sum of signed edge weights. These indices were used to describe the relative connectivity and influence of individual instrumental and relationship dependency items within the network.

**Test-Retest Reliability:** The temporal stability of the scale was assessed using the test-retest method. In this context, data were collected from the same group of participants (n = 55) at two different time points to calculate the total scale score, as well as the scores for the Instrumental Dependency and Relational Dependency subscales. The relationship between the scores obtained from the first and second administrations was examined using the Pearson correlation coefficient.

The internal consistency of the scales utilized in the external validity analysis was assessed by calculating Cronbach's alpha coefficient for each scale. Items that required reverse coding were appropriately recoded before the analysis. The resulting internal consistency coefficients were then reviewed to determine the suitability of the scales for the external validity analysis.

**External Validity:** To assess the external validity of the scale, Multivariate Multiple Regression Analysis was conducted using variables reported to be associated with dependency structures related to LLMs. In this analysis, ATAI (ATAI Acceptance and ATAI Fear subscales), IA, NFC, and the Perceived Trustworthiness of primary LLM were included as external criteria.

The two subscales of the instrument, Instrumental Dependency and Relational Dependency, were used as dependent variables. Subscale scores were calculated by averaging the relevant items to minimize the effects of differences in the number of items between subscales and to enhance score comparability. The ATAI Acceptance and ATAI Fear subscales were computed based on the total scores of the relevant items, with attention control items excluded from the analysis. IA, NFC, and Perceived Trustworthiness of primary LLM were included in the studies using their total scale scores. In the first stage, the multivariate analysis was performed to examine the simultaneous relationships between the specified external criteria and the Instrumental Dependency and Relational Dependency scores. Wilks' Lambda statistic was used to evaluate the extent to which the predictor variables jointly explained the variance in the two dependent variables. To assess the overall significance of the multivariate model, the dependent variable matrix, comprising Instrumental Dependency and Relational Dependency, was compared between the full model, which included all predictors, and the null model containing only the intercept. This comparison was conducted using Wilks' Lambda statistic, thereby testing the explanatory power of all predictors for the two dependent variables within the overall model.

**RESULTS**

**Descriptive Statistics**

Of the participants, 68.5% were female. The majority had a university degree or higher. The sociodemographic characteristics of the participants are presented in Table 2.

Table 2. Sociodemographic Characteristics of Participants

|  |  | Total (N=387) | |
|---|---|---|---|
|  |  | n | % |
| **Gender** | Female | 265 | % 68.5 |
|  | Male | 122 | % 31.5 |
| **Age** | Mean ± SD | 25.22 ± 7.13 | |
| **Education Level** | Secondary School or Below | 2 | % 0.5 |
|  | High School | 10 | % 2.6 |
|  | University or Above | 375 | % 96.9 |
| **Employment Status** | Full-time employment | 117 | % 30.2 |
|  | Part-time employment | 2 | % 0.5 |
|  | Run my own business | 6 | % 1.6 |
|  | Student | 252 | % 65.1 |
|  | Unemployed | 4 | % 1 |
|  | Housewife | 3 | % 0.8 |
|  | Other | 3 | % 0.8 |
| **Income Level** | Very good / Good | 130 | % 33.6 |
|  | Middle | 212 | % 54.8 |
|  | Bad / Very bad | 45 | % 11.6 |

A total of 55.6% of participants reported having a "middle" level of knowledge about large language models (LLMs). When it comes to AI usage, the majority of participants (83.5%) primarily used ChatGPT, while Claude was the least frequently used main tool (0.5%). Participants used primary LLMs in both stationary settings and while on the move, with 55.3% of them engaging in this manner. Mobile devices were the most common means of access, utilized by 77.8% of participants. Additionally, a significant majority preferred text-based interaction with their primary LLM, accounting for 79.8%.

**Validity and Reliability Analyses**

The mean and standard deviation values for all items in each of the two subscales of the LLM-D12 are presented in Table 3.

Table 3. Descriptive Statistics for the Instrumental and Relationship Dependency Scale Items

| Instrumental Dependency | | | Relational Dependency | | |
|---|---|---|---|---|---|
| Item Code | Mean | SD | Item Code | Mean | SD |
| Without_it_less_confident | 1.87 | 1.06 | Share_private_life | 2.38 | 1.43 |
| Immersed_in_it | 2.18 | 1.24 | Genuine_companion | 2.3 | 1.42 |
| More_rewarded | 2.99 | 1.32 | Feel_less_alone | 2.090 | 1.35 |
| Turn_to_it_even_when_able | 2.64 | 1.39 | It_adds_to_my_social_life | 2.02 | 1.21 |
| It_is_my_go_to | 2.69 | 1.41 | Solely as a tool (reverse) | 3.16 | 1.81 |
| Decisions_without_uneasy | 2.03 | 1.16 | No_need_to_talk_to_others | 2.00 | 1.28 |

Monotonicity was evaluated for all 12 items using nonparametric item response analysis with four total-score groups per item (minimum group size = 77). Across items, mean item scores increased with higher total scores, supporting the expected ordering of response probabilities.

For the Instrumental Dependency items, mean scores increased from approximately 0.17-1.16 in the lowest score groups to 1.36-2.68 in the highest score groups. Across these items, the probability of endorsing at least one response category increased from 0.15-0.62 in the lowest groups to 0.72-0.93 in the highest groups, while the probability of endorsing higher response categories (≥3 or ≥4, depending on the item) increased from near zero to 0.23-0.57. No departures from monotonicity were observed for any instrumental item.

For five of the six Relational Dependency items, a similar pattern was observed. Mean item scores increased from 0.04-0.45 in the lowest score groups to 1.83-2.33 in the highest score groups. Correspondingly, the probability of endorsing at least one category increased from 0.04-0.23 to 0.81-0.90, and the probability of endorsing higher categories (≥3) increased from 0.00–0.08 to 0.31-0.48 across groups. These results indicate consistent monotonic increases in item responses for the majority of relational items.

In contrast, the reversed relational item ('using solely as a tool') showed weaker monotonicity. Although mean scores increased modestly across score groups (from 1.90-2.19 to 2.47), higher-category endorsement probabilities did not increase consistently, with the probability of endorsing the highest category decreasing from 0.30 in the lowest group to 0.09 in the highest group. This item also showed markedly lower scalability (H = 0.06) compared with the remaining relational items (H range = 0.36-0.45). Overall, the results indicate that this item may be a candidate for exclusion from the relational dependency scale (Figure 1).

An analysis of changes in Cronbach's alpha upon the removal of specific items showed that the alpha coefficient increased to 0.876 when the item "Solely_as_a_tool (reverse)" was excluded. When including all items, the Cronbach's alpha coefficient for the Relational Dependency subscale was 0.807.

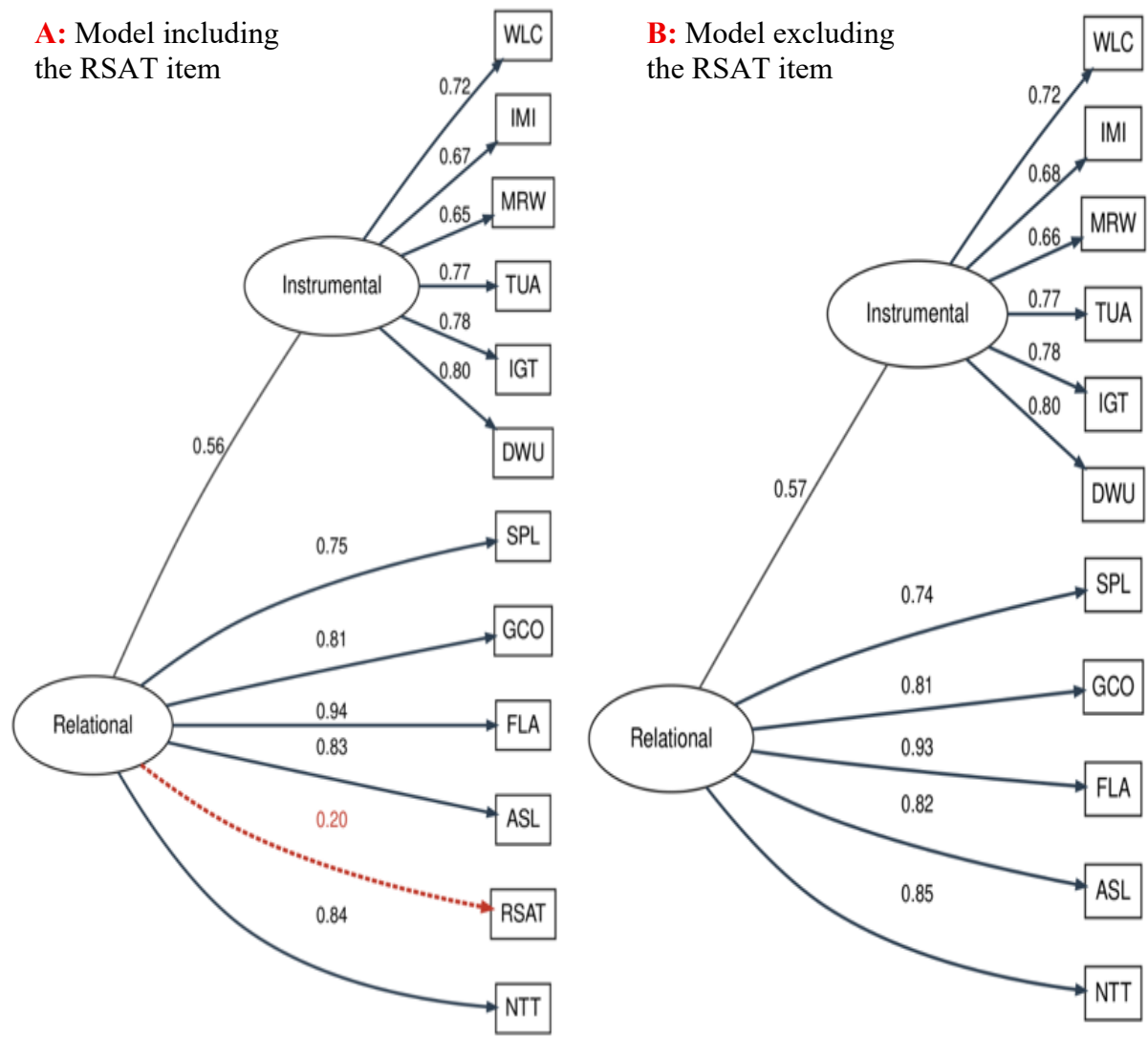

Figure 1. Confirmatory Factor Analysis (CFA) Models of Instrumental and Relational Dependency

**WLC:** Without_it_less_confident. **IMI:** Immersed_in_it. **MRW:** More_rewarded.

**TUA:** Turn to it even when able. **IGT:** It is my go-to. **DWU:** Decisions_without_uneasy

**SPL:** Share_private_life. **GCO:** Genuine_companion. **FLA:** Feel_less_alone.

**ASL:** It adds to my social life. **RSAT:** Solely as a tool (reverse) **NTT:** No need to talk to others

After addressing the problematic item, the remaining items in the Relational Dependency subscale were assessed for their suitability for factor analysis. The KMO value was 0.84, and Bartlett's test of sphericity was significant ($\chi^2 = 1018.56$, df = 10, $p < 0.001$).

In the confirmatory factor analysis (CFA), the standardized factor loadings for the items in the Instrumental Dependency subscale ranged from 0.66 to 0.80. In contrast, those for the Relational Dependency subscale ranged from 0.74 to 0.93. All calculated z-statistics for the items were statistically significant ($p < 0.001$). The explained variance ($R^2$) values for the items ranged from 0.431 to 0.638 in the Instrumental Dependency subscale and from 0.552 to 0.874 in the Relational Dependency subscale. Detailed findings at the item level, including uniqueness values and standard error estimates, are presented in Table 4.

Table 4. Standardized factor loadings, standard errors (SE), uniqueness, and explained variance ($R^2$) for items of the Instrumental and Relational Dependency

| Scale | Item Code | Loading | SE | z (p < 0.001) | Uniqueness | $R^2$ |
|---|---|---|---|---|---|---|
| **Instrumental Dependency** | Without_it_less_confident | 0.72 | 0.030 | 23.952 | 0.488 | 0.512 |
| | Immersed_in_it | 0.68 | 0.032 | 20.849 | 0.543 | 0.457 |
| | More_rewarded | **0.66** | 0.031 | 21.393 | 0.569 | **0.431** |
| | Turn_to_it_even_when_able | 0.77 | 0.026 | 30.166 | 0.400 | 0.600 |
| | It_is_my_go_to | 0.78 | 0.024 | 32.035 | 0.394 | 0.606 |
| | Decisions_without_uneasy | **0.80** | 0.025 | 32.284 | 0.362 | **0.638** |
| **Relational Dependency** | Share_private_life | **0.74** | 0.027 | 27.274 | 0.448 | **0.552** |
| | Genuine_companion | 0.81 | 0.021 | 38.791 | 0.344 | 0.656 |
| | Feel_less_alone | **0.93** | 0.013 | 72.931 | 0.126 | **0.874** |
| | It_adds_to_my_social_life | 0.82 | 0.019 | 43.117 | 0.320 | 0.680 |
| | no_need_to_talk_to_others | 0.85 | 0.021 | 39.554 | 0.285 | 0.715 |

In the two-factor total scale model, which includes the Instrumental Dependency and Relational Dependency subscales, the chi-square value was $\chi^2 = 131.31$, with 43 degrees of freedom (df), resulting in a $\chi^2/df$ ratio of 3.05. The Comparative Fit Index (CFI) for the model was 0.993, the Tucker-Lewis Index (TLI) was 0.991, and the Root Mean Square Error of Approximation (RMSEA) was 0.073, with the lower bound of the 90% confidence interval for RMSEA calculated as 0.059. The Standardized Root Mean Square Residual (SRMR) value for the model was 0.055.

In the single-factor confirmatory factor analysis (CFA) conducted for the Instrumental Dependency subscale, the chi-square value was $\chi^2 = 41.44$, with degrees of freedom df = 9, resulting in $\chi^2/df = 4.60$. For this model, the CFI was 0.991, the TLI was 0.985, and the RMSEA was 0.097, with the lower bound of the 90% confidence interval for RMSEA calculated as 0.068. The SRMR value was 0.050.

In the single-factor CFA of the Relational Dependency subscale, the chi-square value was $\chi^2 = 23.39$, with degrees of freedom df = 5, resulting in $\chi^2/df = 4.68$. For this model, the CFI was 0.997, the TLI was 0.995, and the RMSEA was 0.098, with a lower bound of 0.060 for the 90% confidence interval of RMSEA. The SRMR value was 0.040.

The estimated network revealed a dense pattern of positive associations within both the instrumental and relationship dependency dimensions, alongside fewer and generally weaker cross-dimensional associations. Most edges retained after regularisation were positive, with a small number of negative associations (Figure 2).

Figure 2. Network structure of instrumental and relationship dependency items

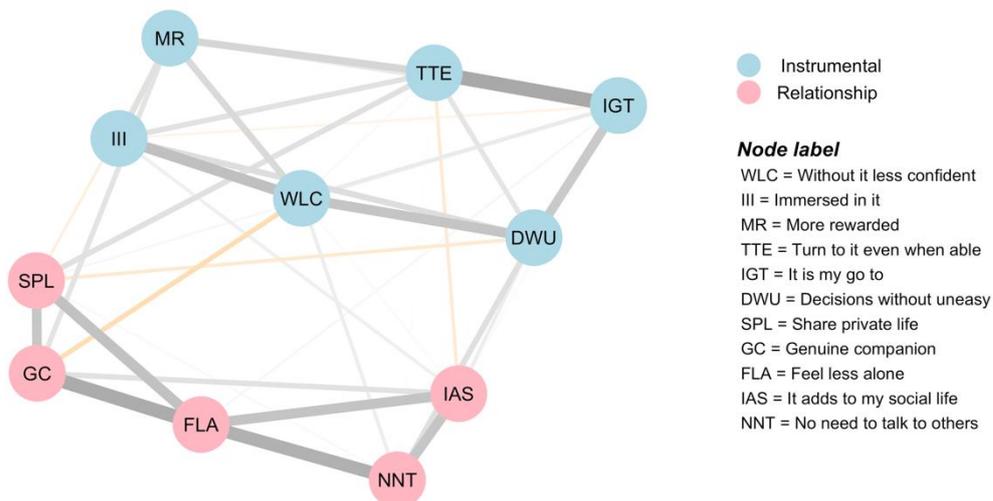

*Note*. Nodes represent individual dependency items. Blue nodes indicate instrumental dependency items, and pink nodes indicate relationship dependency items. Edges represent regularised partial correlations between items, with grey edges indicating positive associations and orange edges indicating negative associations. Edge thickness reflects the magnitude of the association, with thicker edges representing stronger relationships

Within the instrumental dependency, the strongest association was between TTE (Turn to it even when able) and IGT (It is my go to) (weight = 0.41), followed by associations between WLC (Without it less confident) and III (Immersed in it) (weight = 0.30), WLC and DWU (Decisions without uneasy) (weight = 0.28), and IGT and DWU (weight = 0.26). Additional positive associations were observed among other instrumental items, indicating substantial interconnectedness within this dimension.

Within the relationship dependency, the largest edges were observed between GC (Genuine companion) and FLA (Feel less alone) (weight = 0.40), between FLA and NNT (No need to talk to others) (weight = 0.37), and between FLA and IAS (It adds to my social life) (weight = 0.28). SPL (Share private life) was positively associated with both GC (weight = 0.27) and FLA (weight = 0.29), suggesting close coupling among relationship-focused experiences.

Cross-dimensional edges between instrumental and relationship dependency items were generally smaller in magnitude. Positive associations were observed between TTE and SPL (weight = 0.14), and between DWU and NNT (weight = 0.15). A small number of negative cross-dimensional associations were also identified, including between WLC and GC (weight = -0.12), MR (More rewarded) and SPL (weight = -0.05), and DWU and SPL (weight = -0.08). These negative associations were modest relative to the strongest within-dimension edges.

**Validation**

**Internal Consistency**

*The Cronbach's alpha coefficient* was 0.831 for the Instrumental Dependency subscale and 0.876 for the Relational Dependency subscale. When items from both subscales were evaluated together, the Cronbach's alpha coefficient for the total scale was 0.868. When items were removed from the Instrumental Dependency subscale, the resulting Cronbach's alpha coefficients ranged from 0.793 to 0.813, whereas for the Relational Dependency subscale, these values ranged from 0.816 to 0.867.

The corrected item–total correlations (r.cor) ranged from 0.613 to 0.716 for the Instrumental Dependency subscale and from 0.683 to 0.898 for the Relational Dependency subscale. Item–item correlation variance (var.r) values were low in both subscales. The adjusted item–total correlations (r.drop) obtained following item removal are presented in the table. These findings related to internal consistency and item analysis are shown in Table 5.

Table 5. Internal consistency and item analysis indices for Instrumental Dependency and Relational Dependency

| Scale | Item Code | Cronbach's Alpha | alpha SE | r.cor | var.r | r.drop |
|---|---|---|---|---|---|---|
| **Instrumental Dependency** | Without_it_less_confident | 0.807 | 0.015 | 0.661 | 0.007 | 0.596 |
| | Immersed_in_it | **0.813** | 0.015 | 0.625 | 0.005 | 0.555 |
| | More_rewarded | 0.812 | 0.015 | **0.613** | 0.007 | 0.565 |
| | Turn_to_it_even_when_able | **0.793** | 0.017 | **0.716** | 0.004 | 0.654 |
| | It_is_my_go_to | 0.797 | 0.016 | 0.707 | 0.002 | 0.637 |
| | Decisions_without_uneasy | 0.800 | 0.016 | 0.691 | 0.006 | 0.627 |
| **Relational Dependency** | Share_private_life | **0.867** | 0.011 | **0.683** | 0.007 | 0.640 |
| | Genuine_companion | 0.847 | 0.013 | 0.769 | 0.011 | 0.716 |
| | Feel_less_alone | **0.816** | 0.015 | **0.898** | 0.005 | 0.841 |
| | It_adds_to_my_social_life | 0.856 | 0.012 | 0.733 | 0.010 | 0.680 |
| | no_need_to_talk_to_others | 0.859 | 0.012 | 0.726 | 0.009 | 0.665 |

When examining McDonald's omega coefficients for the main scale, it was found that the total omega coefficient (ω) for the Instrumental Dependence subscale was 0.834. The omega total coefficient (ω) for the Relational Dependency subscale was calculated to be 0.880. The omega total coefficient (ω) for the overall scale, which includes both the Instrumental and Relational Dependency subscales, was 0.900.

*Test-retest reliability* was evaluated using total scale scores from the first and second administrations of data collected from 55 participants. A statistically significant correlation was observed between the first and second administrations for the total scale scores (r = 0.697, p < 0.001), with a 95% confidence interval for the correlation coefficient ranging from 0.53 to 0.81. A significant correlation was also noted between the scores for the Instrumental Dependency subscale from the first and second administrations (r = 0.70, p < 0.001), with a 95% confidence interval of 0.54 to 0.81. Similarly, a statistically significant correlation was found for the Relational Dependency subscale scores from the two administrations (r = 0.68, p < 0.001), with a 95% confidence interval of 0.51 to 0.80. These findings indicate that both the total scale score and subscale scores demonstrated consistency across the two administrations [23].

To assess external validity, the internal consistency of the scales used in the study was evaluated using Cronbach's alpha coefficient. The Cronbach's alpha coefficients were 0.608 for the ATAI Acceptance subscale and 0.624 for the ATAI Fear subscale. The coefficient for the IA was 0.836. After appropriate reverse coding of items in the NFC, the Cronbach's alpha coefficient was found to be 0.622. The Cronbach's alpha coefficient for the PT-LLM-08 scale was 0.885, indicating a high level of internal consistency.

**External Validity**

To assess external validity, the relationships between the ATAI Acceptance, ATAI Fear, IA, NFC, and PT-LLM-08 scales and the scores for Instrumental Dependency and Relational Dependency were examined. A comparison of the null model, which contained only the constant term, with the complete model, which included all predictor variables, indicated that the overall multivariate model was statistically significant (Wilks' $\Lambda = 0.826$, approximate $F(10, 760) = 9.61$, $p < 0.001$).

According to Wilks' Lambda test, the results showed statistically significant relationships for several variables. Specifically, the ATAI Acceptance (Wilks' $\Lambda = 0.947$, $F(2, 380) = 10.63$, $p < 0.001$), ATAI Fear (Wilks' $\Lambda = 0.958$, $F(2, 380) = 8.32$, $p < 0.001$), IA (Wilks' $\Lambda = 0.907$, $F(2, 380) = 19.39$, $p < 0.001$), and the PTLLM scale (Wilks' $\Lambda = 0.953$, $F(2, 380) = 9.43$, $p < 0.001$) were all statistically significant with respect to the two dependent variables. In contrast, the NFC variable did not show a significant relationship within the overall multivariate model (Wilks' $\Lambda = 0.988$, $F(2, 380) = 2.30$, $p = 0.102$).

**DISCUSSION**

Based on the findings of this study, the LLM-D12, which consists of two subscales, is deemed a valid and reliable tool for assessing LLM dependency in the general population of Turkish society. The final Turkish version of the LLM-D12 includes the Instrumental Dependency subscale, comprising six items, and the Relational Dependency subscale, consisting of five items. Both subscales have demonstrated validity and reliability. The original version of the scale was developed for the UK population, where its validity was confirmed [5].

Additionally, the item-total correlation coefficient is expected to be non-negative and preferably above +0.25; the value for the RSAT item is 0.21. It is recommended that items failing to meet this criterion be excluded from the scale. Moreover, an increase in the Cronbach's alpha coefficient when an item is removed, compared to the alpha coefficient calculated for the entire scale, indicates that the item may reduce the internal consistency of the scale. In our study, the

Cronbach's alpha coefficient for the Relational Dependency subscale was 0.807 when all items were included, whereas it increased to 0.876 upon exclusion of the RSAT item [3]. Considering all these findings together, it was observed that removing the RSAT item strengthened the measurement model, and the item was subsequently excluded. The exclusion of this reverse-coded item aligns with the 'method effect' phenomenon frequently cited in psychometric literature, where negative wording often disrupts the factor structure due to cognitive processing artifacts rather than the actual trait being measured. Furthermore, as LLMs evolve into sophisticated conversational agents, the conceptual boundary between viewing them 'solely as a tool' and perceiving them as 'interactive partners' becomes increasingly blurred. Consequently, the low loading of this item may reflect the participants' genuine difficulty in compartmentalizing the pragmatic and social aspects of LLM usage, rather than a mere measurement error.

The network analysis provides complementary support for the distinction between instrumental and relationship dependency by showing that items cluster primarily within their respective scales, while remaining interconnected. At the same time, the presence of selective cross-domain connections demonstrates how instrumental reliance and relationship dependencies may co-occur in practice without collapsing into a single undifferentiated construct. From an applied perspective, this structure is informative for practitioners, as it suggests that instrumental and relationship dependency may require different points of emphasis in assessment and intervention. Identifying highly connected items within each type of dependency can help clinicians and practitioners focus on specific experiences that may maintain dependency patterns, while the limited cross-domain links indicate where interventions targeting one domain may have broader downstream effects. Moreover, a detailed mapping of individual items to their underlying theoretical constructs, along with a guideline on the feasibility of using these items as micro-measures, is provided in Appendix 2.

In the evaluation of model fit, the two-factor measurement model demonstrated acceptable fit indices with an RMSEA of 0.073, falling well below the recommended threshold of 0.08. For the single-factor subscale analyses, RMSEA values were observed at 0.097 (Instrumental Dependency) and 0.098 (Relational Dependency). While these values approach the upper limit of acceptability, they must be interpreted in the context of model parsimony. The subscale models possess extremely low degrees of freedom (df = 9 and df = 5, respectively). Methodological research highlights that RMSEA tends to be artificially inflated in models with small degrees of freedom, often rejecting valid models erroneously [24]. Given that the CFI and

TLI values for these subscales were exceptionally high (≥ 0.985) and SRMR values were low (≤ 0.050), the elevated RMSEA values are interpreted as statistical artifacts of low degrees of freedom rather than indicators of model misspecification [25, 26].

In the study, the Cronbach's alpha coefficient was found to be 0.868. According to the literature, a Cronbach's alpha value above 0.70 indicates that the measurement tool is of acceptable quality and that internal consistency is adequately ensured [27].

The test–retest results of this study indicate that the LLM-D12 produces consistent measurements over time. The presence of moderate-to high and statistically significant correlations between the first and second administrations for the total scale score and both subscale scores demonstrates that the measurement stability of the scale is adequate [26, 27].

The limitations of correlation analyses in assessing external validity are well-documented in the literature, and regression-based approaches are recommended as an alternative [26]. Accordingly, multivariate multiple regression, which allows simultaneous evaluation of the two sub-dimensions of the scale, was employed in this study. The finding that the multivariate model was generally significant indicates that the instrumental and relational dependency dimensions of the scale exhibit relationships consistent with theoretically expected patterns. Furthermore, the identification of significant associations between ATAI Acceptance, ATAI Fear, IA, and PT-LLM-08, as well as both sub-dimensions, supports the external validity of the scale. In contrast, no significant relationship was observed for the NFC variable within the model. Contrary to expectations, no significant relationship was observed for the Need for Cognition (NFC) variable. This non-significant finding is particularly revealing; it challenges the common assumption that AI dependency is primarily driven by 'cognitive miserliness' or a desire to avoid thinking. Instead, it suggests that LLM dependency is ubiquitous across cognitive styles. Individuals with high need for cognition may be equally prone to developing instrumental dependency, utilizing LLMs not to escape thinking, but for 'cognitive offloading' to enhance productivity and manage complex information processing. Thus, the risk of dependency appears to transcend intellectual disposition, affecting both those who seek to reduce mental effort and those who seek to augment their cognitive capabilities.

Behavioral addiction theory posits that non-substance addictions encompass specific components such as salience, mood modification, tolerance, withdrawal, conflict, and relapse. While the LLM-D12 aligns with the behavioral addiction framework, it is specifically designed to assess 'dependency' a construct characterizing high-level functional and emotional reliance rather than a clinical addiction diagnosis. The scale's items robustly capture core features of problematic involvement, including increasing dominance of use (salience), reliance for emotional regulation (mood modification), and uneasiness when access is restricted (withdrawal/conflict). However, unlike clinical addiction tools that strictly require the presence of 'relapse' and total loss of control, the LLM-D12 highlights the maladaptive attachment and habitual reliance that may serve as precursors to pathological addiction. Thus, the findings suggest that instrumental and relational dependency on LLMs represents a distinct spectrum of problematic use that shares theoretical roots with addiction but functions primarily through cognitive offloading and para-social bonding [28].

*Limitations*: This study encompasses several limitations that should be considered. First, regarding sample characteristics, while the participants exhibited diversity in age and gender, the majority held a university degree or higher. Therefore, while the findings may not fully represent the general Turkish population, they provide valid insights into highly educated groups who are often the early adopters and primary users of AI technologies. second, given the rapidly evolving capabilities of LLMs, user dependency patterns are dynamic; thus, the scale's validity should be periodically reassessed as the technology matures.

Finally, regarding the theoretical framework, the LLM-D12 was intentionally designed to assess instrumental and relational dependency dynamics rather than to diagnose clinical behavioral addiction. Consequently, while the scale successfully captures core psychological components such as salience, mood modification, tolerance, and withdrawal, it deliberately excludes a 'relapse' component. This exclusion reflects the instrument's specific focus on functional reliance and emotional attachment in human-AI interaction, distinguishing it from pathological addiction models characterized by loss of control. Therefore, researchers aiming to diagnose clinical pathology should view this scale as a measure of dependency intensity, potentially supplementing it with clinical criteria for diagnostic purposes.

**CONCLUSION**

This study thoroughly assessed the validity and reliability of the Turkish adaptation of the LLM-D12 scale, making a significant contribution to the measurement of dependency tendencies toward large language models (LLMs). Originally developed as a 12-item instrument by Yankouskaya et al.[5], the scale was restructured into an 11-item format during the Turkish adaptation process to improve its psychometric coherence.

The LLM-D12 focuses on general dependency tendencies toward LLMs rather than targeting a specific model, thereby addressing a crucial gap in the literature. This enables the assessment of rapidly emerging behaviors associated with LLM usage. Additionally, the lack of a competing Turkish measurement tool in this area underscores the originality of the study at both national and academic levels.

The findings support the two-factor structure of the scale within the Turkish context, and the instrument demonstrates reliable performance. However, re-evaluating the scale in future research across diverse sociodemographic groups and over time will further expand its potential applications.

**Declarations**

**Author Contributions:** Conceptualization, T.C.A., H.D., A.B., and R.A.; methodology, T.C.A., H.D., A.Y., and R.A.; software/formal analysis, H.D., A.Y., and T.C.A.; investigation/data collection, T.C.A. and G.U.; resources, A.B. and H.D.; data curation, H.D., S.A and A.Y.; writing—original draft preparation, T.C.A., G.U., A.Y., and A.B.; writing— review and editing, H.D., Y.K., S.A., A.Y., and R.A.; visualization, H.D. and A.Y.; supervision, H.D., A.B., and R.A. All authors have read and agreed to the published version of the manuscript.

**Funding**: This publication was supported by NPRP 14 Cluster grant # NPRP 14C-0916–210015 from the Qatar National Research Fund (a member of Qatar Foundation). The findings herein reflect the work and are solely the responsibility of the authors.

**Institutional Review Board Statement:** The study was conducted in accordance with the Declaration of Helsinki and approved by the Ethics Committee of Erciyes University Health Sciences Research Ethics Committee (Decision No. 2025/390, Date: August 6, 2025).

**Informed Consent Statement**: Informed consent was obtained from all subjects involved in the study.

**Data Availability Statement**: The datasets generated and/or analyzed during the current study, along with the R scripts used for Monotonicity and Network analyses and SPSS output files, are openly available in the Open Science Framework (OSF) repository: https://osf.io/cs5rw.

**Conflicts of Interest:** The authors declare no conflicts of interest.

# Appendix 1

## Large Language Model Dependency Questionnaire (LLM-D12) – The English Version

Thinking specifically about the large language model (LLM) you use most often (such as ChatGPT, DeepSeek, Gemini, Copilot, etc.), please indicate the extent to which you agree or disagree with each of the following statements:

| Item nr | item | Strongly Disagree | Disagree | Somewhat Disagree | Somewhat Agree | Agree | Strongly Agree |
|---|---|---|---|---|---|---|---|
| ID1 | Without it, I feel less confident when making decisions | | | | | | |
| ID2 | I use it sometimes without realizing how much time I spend immersed in it | | | | | | |
| ID3 | I feel much more rewarded and pleased when completing tasks using it | | | | | | |
| ID4 | I turn to it for support in decisions, even when I can make them myself with some effort | | | | | | |
| ID5 | It is my go-to for assistance in decision-making | | | | | | |
| ID6 | Making decisions without it feels somewhat uneasy | | | | | | |
| RD1 | I share details about my private life with it | | | | | | |
| RD2 | I interact with it as if it were a genuine companion | | | | | | |
| RD3 | It helps me feel less alone when I need to talk to someone | | | | | | |
| RD4 | It adds to my social life, making socializing more engaging and interesting | | | | | | |
| RD-5 [R] | I use it solely as a tool, not to express my feelings or expect it to understand me | | | | | | |
| RD6 | It helps me feel less alone, reducing the need to talk to others | | | | | | |

**Scoring:** Items are rated on a 6-point scale: 1 = Strongly Disagree, 2 = Disagree, 3 = Somewhat Disagree, 4 = Somewhat Agree, 5 = Agree, 6 = Strongly Agree.
[R] indicates that the item shall be reversed when scoring
ID – Instrumental Dependency, RD – Relationship Dependency
Add up the item scores for ID to get the Instrumental Dependency score.
Add up the item scores for RD to get the Relationship Dependency score.
The overall LLM Dependency score is the sum of the two subscale scores.
**Notes:** The LLM examples in the introductory context can be updated to reflect either the most popular models at the time or those most relevant to the context in which the scale is administered.

# The Turkish – Büyük Dil Modelleri Bağımlılık Ölçeği (LLM-D12-TR)

En sık kullandığınız büyük dil modelini (BDM) (örneğin ChatGPT, DeepSeek, Gemini, Copilot vb.) özellikle göz önünde bulundurarak, aşağıdaki ifadelere ne ölçüde katıldığınızı ya da katılmadığınızı belirtiniz:

| Madde No | Madde | Kesinlikle Katılmıyorum | Katılmıyorum | Kısmen Katılmıyorum | Kısmen Katılıyorum | Katılıyorum | Kesinlikle Katılıyorum |
|---|---|---|---|---|---|---|---|
| AB1 | O olmadan karar verirken kendime olan güvenim azalıyor. | | | | | | |
| AB2 | Ne kadar zaman geçirdiğimin farkına varmadan bazen onu kullanıyorum. | | | | | | |
| AB3 | Görevleri onunla tamamladığımda çok daha tatmin olmuş ve memnun hissediyorum. | | | | | | |
| AB4 | Biraz çaba göstererek kendim de verebileceğim kararlar için bile ona başvurduğum oluyor. | | | | | | |
| AB5 | Karar verme süreçlerinde ilk başvurduğum destek kaynağı odur. | | | | | | |
| AB6 | Onu kullanmadan karar vermek beni bir miktar huzursuz hissettiriyor. | | | | | | |
| İB1 | Özel hayatıma dair ayrıntıları onunla paylaşıyorum. | | | | | | |
| İB2 | Onunla etkileşim kurarken gerçek bir arkadaşmış gibi davranıyorum. | | | | | | |
| İB3 | Konuşacak birine ihtiyaç duyduğumda, kendimi daha az yalnız hissetmemi sağlıyor. | | | | | | |
| İB4 | Sosyal hayatıma katkıda bulunuyor sosyalleşmeyi daha ilgi çekici ve keyifli hâle getiriyor. | | | | | | |
| İB5 | Onunla etkileşim kurmak, başkalarıyla konuşma ihtiyacımı azaltarak kendimi daha az yalnız hissetmemi sağlıyor. | | | | | | |

**Puanlama:** Maddeler 6'lı bir ölçek üzerinden değerlendirilir: 1 = Kesinlikle Katılmıyorum, 2 = Katılmıyorum, 3 = Kısmen Katılmıyorum, 4 = Kısmen Katılıyorum, 5 = Katılıyorum, 6 = Kesinlikle Katılıyorum.
AB – Araçsal Bağımlılık, İB – İlişkisel Bağımlılık
İlk 6 madde araçsal bağımlılık alt boyutunu, sonraki 5 soru ilişkisel bağımlılık alt boyutunu ölçmektedir.
Ters madde bulunmamaktadır.
Araçsal Bağımlılık puanını elde etmek için AB maddelerinin puanlarını toplayın.
İlişkisel Bağımlılık puanını elde etmek için İB maddelerinin puanlarını toplayın.
Genel BDM Bağımlılığı puanı, bu iki alt ölçek puanının toplamıdır.
**Notlar:** *Türkçe adaptasyonu sırasında orijinal ölçekten farklı olarak 1 madde (RD-5 [R]: Sadece araç olarak kullanma maddesi) çıkarılmıştır.

Giriş metnindeki BDM örnekleri, o anki en popüler modelleri veya ölçeğin uygulandığı bağlama en uygun olanları yansıtacak şekilde güncellenebilir.

**Notes:** The Turkish adaptation of the LLM-D12 scale presented here is based on the original structure developed by Yankouskaya et al. (2025) [5]. For cross-cultural validation context, see also the Arabic validation study by AlShakhsi et al. (2025) [29].

# APPENDIX 2

## Mapping individual items of LLM-D12 into underlying theoretical constructs and measurements

| Item nr | item | Underlying theoretical construct | Measuring the extent of |
|---|---|---|---|
| ID1 | Without it, I feel less confident when making decisions | **Decision-making confidence** | cognitive dependency on the LLM for decision-making confidence |
| ID2 | I use it sometimes without realizing how much time I spend immersed in it | **Immersive engagement** | becoming deeply absorbed in the interaction to the point of reduced awareness of surroundings or schedules |
| ID3 | I feel much more rewarded and pleased when completing tasks using it | **Positive reinforcement from use** | experiencing greater gratification or accomplishment when the LLM is part of the task process |
| ID4 | I turn to it for support in decisions, even when I can make them myself with some effort | **Cognitive offloading** | leaning on the LLM to reduce cognitive effort |
| ID5 | It is my go-to for assistance in decision-making | **Decision-making reliance** | a person turns to the LLM as their primary source of help when making decisions |
| ID6 | Making decisions without it feels somewhat uneasy | **Decision-making unease** | a person experiences discomfort or uncertainty in decision-making when LLM is unavailable (sense of altered decision experience) |
| RD1 | I share details about my private life with it | **Self-disclosure** | a person shares personal or private information with LLM (based on the belief that the interaction is private and non-judgmental) |
| RD2 | I interact with it as if it were a genuine companion | **Companionship** | assigning human-like qualities, intentions, or responsiveness to the LLM in the context of ongoing interaction |
| RD3 | It helps me feel less alone when I need to talk to someone | **Perceived social connectedness** | perceived social connectedness through LLM interaction |
| RD4 | It adds to my social life, making socializing more engaging and interesting | **Positive social enhancement** | belief that one's social life is more engaging, dynamic, or stimulating because of the LLM |
| RD-5 [R] | I use it solely as a tool, not to express my feelings or expect it to understand me | **Withholding personal expression** | one avoids sharing inner states such as opinions, experiences, values, or emotions |
| RD6 | It helps me feel less alone, reducing the need to talk to others | **Social substitution** | LLM interaction replaces interaction with other people because the LLM fulfils that role |

**Note on feasibility of using LLM-D12 at its item level**

Examining responses to individual items in the LLM-D12, rather than relying solely on total subscale scores, can provide a more detailed understanding of the specific behaviours within instrumental and relationship LLM dependency.

This approach may be particularly beneficial when:

- Identifying distinct behaviours that could inform targeted interventions.
- Comparing specific aspects of LLM use between different groups or across measurements.

- Determining which item of dependency are most closely associated with other psychological or behavioural variables.
- Conducting exploratory or formative studies where fine-grained behavioural insights are needed.
- Estimating prevalence of individual indicators of dependent behaviour.
- Identifying the hierarchy of indicators of Instrumental or Relationship LLM dependency.

The psychometric properties of the LLM-D12 indicate that the use of individual items as micromeasures is feasible. Both subscales showed good internal consistency and a coherent factor structure, suggesting that each item is meaningfully related to the construct measured by its subscale. However, the following limitations should be considered:

- Each item contributes differently to the corresponding subscale, so item scores cannot be treated as equivalent.
- The relationships between items within each subscale differ in strength. It is recommended to review item loadings and inter-item relationships, such as through factor analysis or network modelling before interpreting individual item scores.
- Interpreting each item as if it measures a separate construct may be misleading, as items within a subscale share variance due to their focus on related aspects of the same underlying concept.

## LLM-D12'nin Bireysel Maddelerinin Altında Yatan Teorik Yapılar ve Ölçümlerle Eşleştirilmesi

| Madde No | Madde | Altında Yatan Teorik Yapı | Ölçülen Kapsam |
| --- | --- | --- | --- |
| AB1 | O olmadan karar verirken kendime olan güvenim azalıyor. | **Karar Verme Özgüveni** | Karar verme güveni konusunda LLM'e (Büyük Dil Modeli) yönelik bilişsel bağımlılık. |
| AB2 | Ne kadar zaman geçirdiğimin farkına varmadan bazen onu kullanıyorum. | **Sürükleyici Etkileşim** | Çevresel farkındalığın veya zaman algısının azalacağı düzeyde etkileşime derinlemesine dalma. |
| AB3 | Görevleri onunla tamamladığımda çok daha tatmin olmuş ve memnun hissediyorum. | **Kullanımdan Kaynaklı Olumlu Pekiştirme** | Görev sürecine LLM dahil olduğunda daha yüksek bir tatmin veya başarı hissi yaşama. |
| AB4 | Biraz çaba göstererek kendim de verebileceğim kararlar için bile ona başvurduğum oluyor. | **Bilişsel Yükü Hafifletme (Offloading)** | Bilişsel çabayı azaltmak amacıyla LLM'e yaslanma. |
| AB5 | Karar verme süreçlerinde ilk başvurduğum destek kaynağı odur. | **Karar Vermede Dayanak** | Bireyin karar verirken birincil yardım kaynağı olarak LLM'e yönelmesi. |
| AB6 | Onu kullanmadan karar vermek beni bir miktar huzursuz hissettiriyor. | **Karar Verme Huzursuzluğu** | LLM erişilebilir olmadığında karar verme sürecinde rahatsızlık veya belirsizlik yaşama (değişime uğramış karar deneyimi hissi). |
| İB1 | Özel hayatıma dair ayrıntıları onunla paylaşıyorum. | **Kendini Açma (İfşa)** | Bireyin (etkileşimin gizli ve yargılamadan uzak olduğu inancına dayanarak) kişisel veya özel bilgileri LLM ile paylaşması. |
| İB2 | Onunla etkileşim kurarken gerçek bir arkadaşmış gibi davranıyorum. | **Yoldaşlık (Arkadaşlık)** | Devam eden etkileşim bağlamında LLM'e insani nitelikler, niyetler veya tepkisellik atfetme. |
| İB3 | Konuşacak birine ihtiyaç duyduğumda, kendimi daha az yalnız hissetmemi sağlıyor. | **Algılanan Sosyal Bağlılık** | LLM etkileşimi yoluyla algılanan sosyal bağlılık hissi. |
| İB4 | Sosyal hayatıma katkıda bulunuyor sosyalleşmeyi daha ilgi çekici ve keyifli hâle getiriyor. | **Olumlu Sosyal Geliştirme** | LLM sayesinde bireyin sosyal hayatının daha ilgi çekici, dinamik veya uyarıcı hale geldiği inancı. |
| İB5 | Onunla etkileşim kurmak, başkalarıyla konuşma ihtiyacımı azaltarak kendimi daha az yalnız hissetmemi sağlıyor. | **Sosyal İkame** | LLM'in bu rolü üstlenmesi nedeniyle, LLM ile etkileşimin diğer insanlarla olan etkileşimin yerini alması. |

## LLM-D12'nin Madde Düzeyinde Kullanımının Uygulanabilirliğine Dair Not

Yalnızca toplam alt ölçek puanlarına dayanmak yerine, LLM-D12'deki bireysel maddelere verilen yanıtların incelenmesi, Araçsal ve İlişkisel LLM bağımlılığı kapsamındaki spesifik davranışların daha ayrıntılı bir şekilde anlaşılmasını sağlayabilir.

Bu yaklaşım özellikle aşağıdaki durumlarda faydalı olabilir:

- Hedefe yönelik müdahaleleri şekillendirebilecek belirgin davranışların tanımlanması.
- LLM kullanımının belirli yönlerinin farklı gruplar veya ölçümler arasında karşılaştırılması.
- Hangi bağımlılık maddesinin diğer psikolojik veya davranışsal değişkenlerle en yakından ilişkili olduğunun belirlenmesi.
- İnce taneli davranışsal içgörülerin gerekli olduğu keşifsel veya biçimlendirici çalışmaların yürütülmesi.
- Bağımlı davranışa ait bireysel göstergelerin yaygınlığının tahmin edilmesi.
- Araçsal veya İlişkisel LLM bağımlılığı göstergelerinin hiyerarşisinin belirlenmesi.

LLM-D12'nin psikometrik özellikleri, bireysel maddelerin mikro ölçümler olarak kullanılmasının uygulanabilir olduğunu göstermektedir. Her iki alt ölçek de iyi bir iç tutarlılık ve tutarlı bir faktör yapısı sergilemiştir; bu durum, her bir maddenin, ait olduğu alt ölçek tarafından ölçülen yapıyla anlamlı bir ilişki içinde olduğunu düşündürmektedir. Bununla birlikte, aşağıdaki sınırlılıklar dikkate alınmalıdır:

- Her madde ilgili alt ölçeğe farklı şekilde katkıda bulunmaktadır, bu nedenle madde puanları eşdeğer olarak değerlendirilemez.
- Her bir alt ölçekteki maddeler arasındaki ilişkilerin gücü farklılık göstermektedir. Bireysel madde puanlarını yorumlamadan önce, faktör analizi veya ağ modellemesi (network modelling) gibi yöntemlerle madde yüklerinin ve maddeler arası ilişkilerin gözden geçirilmesi önerilir.
- Her bir maddeyi ayrı bir yapıyı ölçüyormuş gibi yorumlamak yanıltıcı olabilir; zira bir alt ölçekteki maddeler, aynı temel kavramın ilişkili yönlerine odaklandıkları için ortak bir varyansı paylaşmaktadır.

**Notes:** The Turkish adaptation of the LLM-D12 scale presented here is based on the original structure developed by Yankouskaya et al. (2025) [5]. For cross-cultural validation context, see also the Arabic validation study by AlShakhsi et al. (2025) [29].